\documentclass{aa}  
\usepackage[colorlinks=true,linkcolor=blue,citecolor    = blue]{hyperref}
\usepackage{graphicx}
\usepackage{txfonts}
\usepackage{float}
\usepackage{rotating,tabularx}
\usepackage{fancyhdr}
\usepackage{multirow}
\usepackage{afterpage}
\usepackage{bigstrut}
\usepackage[table,xcdraw]{xcolor}
\usepackage[flushleft]{threeparttable }
\usepackage{amsmath}
\usepackage{enumitem}
\usepackage{amsfonts,color}
\usepackage{amssymb,float}
\usepackage[utf8]{inputenc}
\usepackage{color}
\usepackage{etoolbox}
\usepackage{appendix}
\usepackage{url}
\usepackage{hypcap}
\usepackage{dirtytalk}
\usepackage{fontawesome}

\usepackage{booktabs}
\usepackage[graphicx]{realboxes}
\usepackage{adjustbox}
\usepackage{multirow}
\usepackage{rotating,tabularx}
\usepackage{afterpage}
\usepackage{tablefootnote}
\usepackage{placeins} 
\usepackage{tikz}

\usepackage{bigstrut}

\definecolor{lime}{HTML}{A6CE39}
\DeclareRobustCommand{\orcidicon}{%
    \begin{tikzpicture}
    \draw[lime, fill=lime] (0,0) 
    circle [radius=0.16] 
    node[white] {{\fontfamily{qag}\selectfont \tiny ID}};
    \draw[white, fill=white] (-0.0625,0.095) 
    circle [radius=0.007];
    \end{tikzpicture}
    \hspace{-2mm}
}

\foreach \x in {A, ..., Z}{%
    \expandafter\xdef\csname orcid\x\endcsname{\noexpand\href{https://orcid.org/\csname orcidauthor\x\endcsname}{\noexpand\orcidicon}}
}
\newcommand{\orcid}[1]{\href{https://orcid.org/#1}{\textcolor[HTML]{A6CE39}{\orcidicon}}}
\usepackage{graphicx}

\usepackage{txfonts}

\newcommand{\gaia}{\textit{Gaia}}

\setcitestyle{notesep={: }}
\defcitealias{2024arXiv240208843C}{Paper I}
\defcitealias{2023A&A...674A..22G}{G23}
\defcitealias{2017EPJWC.15201021C}{C17}

\begin{document} 

\title{Variable stars in Galactic globular clusters\\  II.
~Population II Cepheids\thanks{Tables \ref{Tab:Cepheids_Gc} and \ref{Tab:cepheid_parameters} are available in electronic form at the CDS via anonymous ftp to cdsarc.cds.unistra.fr (130.79.128.5)
or via https://cdsarc.cds.unistra.fr/cgi-bin/qcat?J/A+A/}}
\titlerunning{Population II Cepheids in globular clusters}

   \author{
   Mauricio Cruz Reyes \inst{1}\orcid{0000-0003-2443-173X}          \and Richard I. Anderson\inst{1}\orcid{0000-0001-8089-4419} \and Susmita Das\inst{2,3}\orcid{0000-0003-3679-2428}    }

   \institute{Institute of Physics, Laboratory of Astrophysics, \'Ecole Polytechnique F\'ed\'erale de Lausanne (EPFL), 1290 Versoix, Switzerland  \and Konkoly Observatory, HUN-REN Research Centre for Astronomy and Earth Sciences, Konkoly-Thege Miklós út 15-17, H-1121, Budapest, Hungary \and
 CSFK, MTA Centre of Excellence, Budapest, Konkoly-Thege Miklós út 15-17., H-1121, Hungary
\\
    \email{mauricio.cruzreyes@epfl.ch, richard.anderson@epfl.ch}   }

   \date{Received \today}

  \abstract
  {We identified a sample of 88 bona fide Population II Cepheids (henceforth referred to as Cepheids) and 44 candidates in Galactic globular clusters (GCs). Seventy-eight of the Cepheids in the sample align within $2\,\sigma$ of the period-luminosity relation for Milky Way Type II Cepheids (T2CEPs). Nine align with the period-luminosity relation for fundamental-mode anomalous Cepheids (ACEPs), and only one (BL Boötis) follows the relation for first-overtone ACEPs, as determined from observations of ACEPs in the Large Magellanic Cloud.   For sources in common between our catalog and the OGLE catalog, the classification agrees in 94\% of cases. In comparison, for sources shared between the \gaia\ Specific Object Study (SOS) and OGLE, the agreement is 74\%. In the dense environments of GCs, our analysis shows that the completeness of the \gaia\ catalogs for Cepheids is 64\% for the SOS and 74\% for the classifier of variable stars. We determined the red and blue edges of the instability strip for T2CEPs using linear MESA-RSP models.  We find that the best-fit models, with $M = 0.6\,{\rm M}_\odot$ and Z = 0.0003, are able to fit $90\%$ of the stars in our sample. This percentage is the same for helium abundances $Y = 0.220$ and $0.245$.   Higher values of $Y$ lower this percentage, and the same effect is observed with lower values of $Z$. In the future, combining the sample of T2CEPs with the precise parallaxes obtained from GCs will strengthen the geometric calibration of a distance ladder based on Population II stars.  This will be useful for determining distances within the Milky Way and for cross-checking distances to Local Group galaxies determined through other methods. }

   \maketitle

\section{Introduction}
Type II Cepheids (T2CEPs) are Population II stars with pulsation periods ranging from 1 day to 100 days \citep{2022Univ....8..122B} and masses in the range $0.5 - 0.6 M_{\odot}$. Despite their similar name, they are quite different from classical Cepheids (DCEPs). Classical Cepheids are Population I stars and are typically located in areas of recent star formation, mostly in spiral arms and the Galactic disk. Up to $\sim$10\% of these stars reside in open clusters \citep{2013MNRAS.434.2238A,2023A&A...672A..85C,2024AJ....168...34W}. 
Type II Cepheids follow a different period-luminosity relation than DCEPs, making them less luminous at the same pulsation period. This finding allowed \citet{1956PASP...68....5B} to determine that T2CEPs and DCEPs belong to different stellar populations.  Type II Cepheids are classified into three subgroups according to their pulsation periods:  BL~Herculis (1–4 days),  W~Virginis (4–20 days), and RV Tauri (periods longer than 20 days). A fourth subclass also exists, namely the peculiar W~Vir (pW~Vir) stars, which are bluer and brighter than W~Vir stars \citep{soszynski2008}. The different T2CEP subgroups exhibit different light curve structures and are predicted to be at different evolutionary stages \citep{gingold1985, wallerstein2002, bono2020}. The most recent prediction of the evolutionary status of T2CEPs is presented in the review paper by \citet{bono2024}, who posit that BL~Her and W~Vir stars share a common evolutionary channel and are either post-zero age horizontal branch stars in the double-shell burning stage or post-asymptotic giant branch (AGB) stars in the hydrogen-shell burning stage. RV~Tauri stars are mainly post-AGB stars. Most T2CEPs pulsate exclusively in the fundamental mode \citep{bono1997},  although two first-overtone T2CEPs were recently discovered by \citet{soszynski2019} in the Large Magellanic Cloud (LMC). While T2CEPs are considered to be Population~II stars, recent studies suggest that W~Vir stars may originate from binary systems \citep{groenewegen2017} and RV~Tau stars may have younger and massive progenitors \citep{manick2018}.   Similar to RR Lyrae stars (RRLs), T2CEPs are frequently found in globular clusters (GCs), although they are less abundant than RRLs.  

Anomalous Cepheids (ACEPs) have been observed in both intermediate-age (1-6 Gyr) and old (>10 Gyr) stellar populations \citep{2012A&A...540A.102F}.  To date, five ACEPs have been confirmed as members of Milky Way clusters, with one additional candidate in NGC~6541 \citep{2020AcA....70..101S}. The prototype ACEP, BL~Boötis, is located in NGC~5466 \citep{1976AJ.....81..527Z}. One ACEP has been identified in each of the following clusters:  NGC~6341 \citep{2006MNRAS.370.1979M,2020MNRAS.494.3212Y}, NGC~6304 \citep{2006MmSAI..77..117D,2020AcA....70..101S}, NGC~6388 \citep{2015A&A...573A.103S}, and NGC~7078 \citep{2021ApJ...922...20B}. 

In this paper, the term "Cepheids" is used collectively to refer to T2CEPs and ACEPs, and we note that classifications of the same star can differ between catalogs. When necessary, we address each specific type. In general, identifying Cepheids in GCs is more challenging than identifying RRLs because the former have longer pulsation periods. Currently, the catalog of \citet[hereafter, C17]{2017EPJWC.15201021C} is the most comprehensive catalog of Cepheids in clusters. Christine Clement regularly updates the catalog, and the latest version includes 105 bona fide Cepheids and 25 candidates\footnote{\url{https://www.astro.utoronto.ca/~cclement/cat/listngc.html}}.  In comparison, the optical gravitational lensing experiment (OGLE) has identified 53 T2CEPs in the Small Magellanic Cloud and 285 in the LMC \citep{2018AcA....68...89S}.

Catalogs of variable stars in clusters have a wide range of applications. For example, \citet{2024A&A...685A..41S} recently used the T2CEP period-luminosity relations, calibrated using the geometric distance to the LMC \citep{2019Natur.567..200P}, to estimate the distance to 22 GCs. 
\citet{2024arXiv240904259L} used RRL catalogs to trace the origins of their host clusters, and \citet{2024MNRAS.tmp.2277P} used their own RRL catalog to study the Oosterhoff dichotomy. These catalogs can also be used to test the performance of \gaia\ algorithms in identifying and classifying variables in areas with a high density of stars.  In \citet[hereafter Paper I]{2024arXiv240208843C} our analysis for RRLs revealed that approximately 20\% of RRLs within clusters went undetected by the analyses of the \textit{Gaia} collaboration. The undetected RRLs are mostly, but not exclusively, located in the central regions of clusters.

This is the second paper in a series dedicated to studying variable stars in Galactic GCs. While Paper I focused on identifying RRLs within GCs and examining their population properties, this work mainly focuses on the identification of Cepheids in GCs.  This paper is structured as follows: Section \ref{sec:datasets} describes the sample of Cepheids and GCs used in the analysis. Section \ref{sec:preliminary_membership} presents the results of the membership analysis and discusses the limitations of the method. False detections are removed from the sample in Sect. \ref{sec:processing}, and Sect. \ref{sec:analysis} presents the period-luminosity relations for these stars and discusses the completeness of the \gaia\ catalogs for Cepheids in GCs. Section \ref{sec:models} sets constraints on the metallicity and helium abundance of T2CEPs by comparing the MESA-RSP models (Modules for Experiments in Stellar Astrophysics - Radial Stellar Pulsations) with the observations. Section \ref{sec:conclusions} summarizes the paper and presents the conclusions.

\section{Dataset}\label{sec:datasets}
Our approach is analogous to the membership analysis of RRLs in GCs presented in  \citetalias{2024arXiv240208843C}, which is largely based on the sample of 170 GCs and their cluster members compiled by \citet{2021MNRAS.505.5978V}. \citetalias{2024arXiv240208843C} lists the astrometric parameters adopted in our analysis, while the ellipticities of GCs, essential for constructing the membership prior, were obtained from \citet{2024A&A...689A.232C}, and the reddening values were taken from the catalog of cluster parameters by \citet{harris2010new}.

We compiled our sample of variable stars using five different catalogs. The first, the \gaia\  Specific Object Study (SOS), contains 15,021 Cepheids \citep{2023A&A...674A..17R}. The second catalog, the \gaia\ classifier of variable sources, includes 16,141 Cepheids. The third catalog, \citet[henceforth: G23]{2023A&A...674A..22G}, incorporates 19,482 Cepheids. The fourth catalog combines the OGLE collection of T2CEPs in the Galactic disk \citep{2018AcA....68..315U} and the Galactic bulge \citep{2017AcA....67..297S}. After merging the Cepheids from the  \citetalias{2017EPJWC.15201021C} catalog with the previous ones, we had a total of 21,240 unique Cepheids. It is important to note that all ACEPs observed by \citet{2022AJ....164..191N} and all T2CEPs from \citet{2022AJ....164..154N} are included in the sample.

\section{Membership analysis}\label{sec:preliminary_membership}
In order to identify the Cepheid stars associated with GCs, we followed the same method presented in \citetalias{2024arXiv240208843C}. We assessed the membership probability of each variable star near a GC by comparing its astrometry with that of the cluster. This method involves a hypothesis test, where the initial assumption is that the star belongs to the cluster. The intuitive idea is that when the astrometric parameters are similar, the star is likely a cluster member. Conversely, as similarities decrease, the probability also decreases. It should be noted that imprecise data may prevent us from rejecting the hypothesis. The details of this method have been extensively described in \citetalias{2024arXiv240208843C}. 

\begin{figure}
    \centering
    \includegraphics[scale =  0.55]{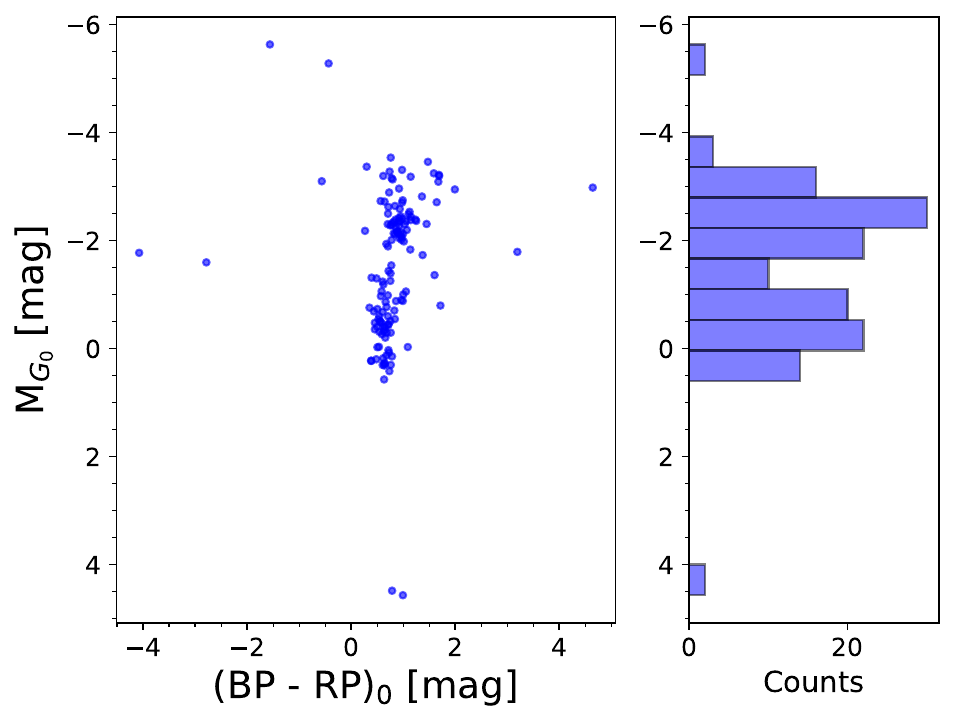}
    \caption{Color-magnitude diagram of the Cepheids in our sample.  The histogram on the left shows the number of stars as a function of absolute magnitude in $G$. Out of the 158 stars in the initial sample, only 141 appear in the plot, as BP or RP photometry is unavailable for 17 of them. }
    \label{fig:cep_cmd}
\end{figure}

Following the membership analysis, we identified an initial sample of 158 Cepheids in clusters. These can be found in Table \ref{Tab:Cepheids_Gc}. However, as can be seen in Fig. \ref{fig:cep_cmd}, some Cepheids are located in areas that are inconsistent with the expected absolute magnitude or intrinsic color for Cepheids, particularly those found below the horizontal branch ($M_{G_{0}} > 0.5$ mag), which is too faint for them. This discrepancy may arise from several factors. For example, large uncertainties in the astrometry of either the cluster or the star can result in high membership probabilities simply because membership cannot be excluded when the constraints are weak. Additionally, some stars might be misclassified.  In particular, W~Vir stars are sometimes misclassified as DCEPs \citep{2023A&A...674A..17R}, and RRLs as ACEPs. A good example is the prototype star of ACEPs, located in the cluster NGC~5466. This star was classified as an RRL in the \citetalias{2023A&A...674A..22G} catalog, which led to its accidental classification as a cluster RRL in \citetalias{2024arXiv240208843C}. Additionally,  as highlighted in \citetalias{2024arXiv240208843C}, the complexity near the centers of GCs can occasionally affect the accuracy of the crossmatch between \gaia\ and various sources in the literature. This could impact the crossmatch of the \citetalias{2023A&A...674A..22G} and \citetalias{2017EPJWC.15201021C} catalogs. These stars are removed from the sample in the next section.

\section{Processing of the sample}\label{sec:processing}
In order to clean the sample of Cepheid stars, outliers from the sample presented in the previous section were excluded using a two-step approach. In the first step, we used the color-magnitude diagram to remove stars with absolute magnitudes or colors atypical for Cepheids. In the second step, we eliminated stars that did not align with the period-luminosity relations. Cepheids that met both selection criteria can be identified in Table \ref{Tab:Cepheids_Gc} using the boolean column "Final." Here, "True" indicates that the stars fulfill these criteria.   The first method cannot be applied to all stars because some do not have a measured color.  This occurs due to crowding, which is more common near the center of the cluster and mainly affects the photometry in the \gaia\ BP and RP bands. This is because the instrument used for obtaining photometry in these bands has a lower resolution than that used in the G band. Similarly, the second method is not applicable to all stars, since not all Cepheids have a measured period. We refer to stars for which none of the methods can be applied as "candidates." These stars can be identified in Table  \ref{Tab:Cepheids_Gc} using the boolean column labeled "Candidate."

For the first step, we selected all stars from the Cepheid sample discussed in the previous section that are brighter than the horizontal branch ($M_{G_{0}} < 0.23 $ mag) and fall within the color range $0 < \mathrm{(BP - RP)}_{0} < 2$ mag. In the second step, we used the \gaia\ Wesenheit magnitude W$\mathrm{_{G} = G - R^{W} (BP - RP)}$, with $R^{W}$ = 1.9 \citep{2019A&A...625A..14R} to visualize the period-luminosity relations. We estimated the absolute magnitude of each star using the GC distances determined by \citet{2021MNRAS.505.5957B}.  For Cepheids, \gaia\ offers two types of photometry: one available in the source catalog and the other in the SOS catalog. On average, the W$_\mathrm{G}$ magnitudes of the stars in the SOS catalog are $0.04$ mag fainter than their counterparts in the source catalog. In the most extreme cases, the apparent magnitudes can differ by as much as $11$ mag. This discrepancy arises because the \gaia\ source catalog does not consider the variability of Cepheids when determining their mean magnitudes, whereas the SOS does (see \citealt{2023A&A...674A..13E} for a detailed explanation). Since not all stars in our sample are included in the SOS, to remove outliers we used SOS photometry when available, and the source catalog otherwise. 

The geometric calibration of period-luminosity relations in clusters is complex for several reasons. The parallaxes of some clusters have  relatively high uncertainties, which complicates the process of estimating their distances. Moreover, when multiple Cepheids are found within the same cluster, estimating their distances using the cluster parallax leads to correlated absolute magnitudes among these stars. Ignoring these correlations can lead to an underestimation of the uncertainties associated with the period-luminosity relation.  Thus, in this study we decided to use the period-luminosity relations previously determined in the literature, and leave the calibration for a separate paper (Lengen et al., in prep.). In particular, we used the period-luminosity relations determined by \citet{2019A&A...625A..14R} for field Milky Way T2CEPs and ACEPs located in the LMC. To convert from apparent to absolute magnitudes, we used the LMC distance determined by \citet{2019Natur.567..200P}.   For each Cepheid, we computed the total uncertainty, $\sigma$, as the sum in quadrature of the uncertainties in photometry and the uncertainties associated with the parameters of the period-luminosity relation. In the case of ACEPs, we further included the uncertainty of the LMC distance. We considered stars that fall within $2\,\sigma$ of a given period-luminosity relation as belonging to that sequence.  If a star satisfies this criterion for more than one period-luminosity relation, we selected the one for which the difference is smallest.  All sources without SOS photometry, rejected using the period-luminosity relations and simultaneously listed in the \citetalias{2017EPJWC.15201021C} catalog, remained as candidates. A total of 20 sources meet this criterion. There are also 24 additional sources that are part of this group because they do not have a measured color or period.  The star Gaia~DR3~4048895253682114432 aligns with the period-luminosity relation of T2CEPs, but it was removed from the sample because its pulsation period ($P \approx 0.49$ days) is too short for a T2CEP. 

After cleaning the sample, we were left with 88 Cepheids distributed across 39 GCs. No Cepheids were found in 131 of the GCs listed in Sect. \ref{sec:datasets}. This means that around 23\% of the GCs in the Milky Way contain T2CEPs, which contrasts strongly with the result for RRLs, as 68\% of the GCs host them \citepalias{2024arXiv240208843C}.  

\section{Analysis of the sample}\label{sec:analysis}
Seventy-eight stars from the sample align with the period-luminosity relations of T2CEPs. Based on their pulsation periods, 28 of them are BL~Her, 39 are W~Vir, and 11 are RV~Tau. Nine stars align with the period-luminosity relation for fundamental-mode ACEPs, and one star aligns with the period-luminosity relation for first-overtone ACEPs.     RV Tau stars represent 19\% of all T2CEPs in the LMC, and they account for 14\% in GCs.  Similarly, W~Vir stars with periods between 4 and 8 days make up 16\% of the T2CEPs in the LMC, but only 9\% in GCs.  

We used the OGLE catalog as a reference to verify the accuracy of our classification. There are 33 sources in common between both catalogs, and our classification agrees for 31 of them, an agreement rate of 94\%. When comparing the OGLE catalog with the SOS from \gaia, there are 23 sources in common, and the classification agrees for 17 of them, a 74\% agreement rate. This is particularly interesting because the estimated contamination of the SOS catalog \citep{2023A&A...674A..13E} across the entire sky is around $1$\%, indicating that misclassifications are more common in regions with a high density of sources. 

Seventeen stars from our sample of Cepheids were classified as DCEPs by the \gaia\ SOS; however, 16 of them follow the period-luminosity relation of T2CEP stars, and one follows the relation of first-overtone ACEPs,  indicating misclassification. For the sources in common, the OGLE and C17 catalogs agree with our classification. The only exception is V12 in NGC~6333, which is classified as an ACEP in the C17 catalog and as a T2CEP by both our analysis and the OGLE Collaboration.

\begin{figure*}
    \centering
    \includegraphics[scale=0.73]{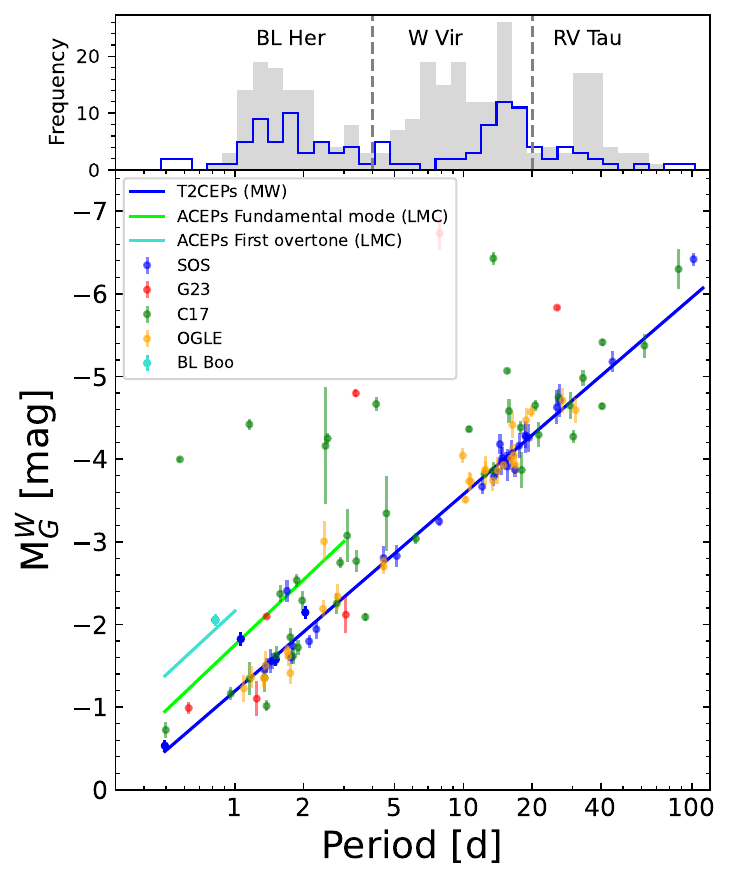}
    \includegraphics[scale=0.73]{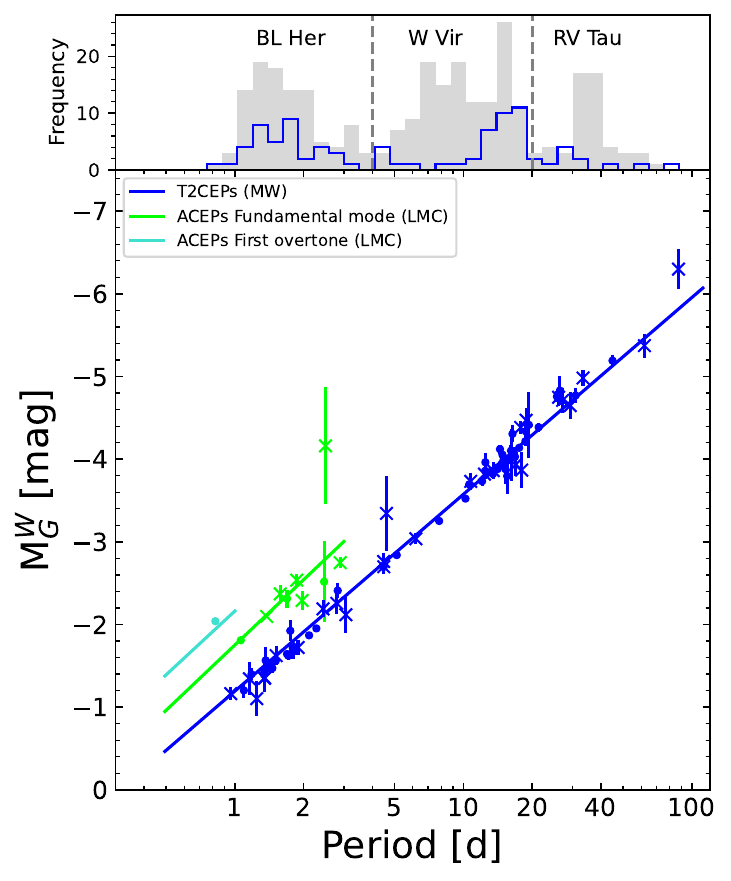}
    \caption{Absolute magnitude of Cepheids in GCs in the \gaia\ Wesenheit magnitude compared with the period-luminosity relations derived by \citet{2019A&A...625A..14R}. The period-luminosity relation for T2CEPs is shown in blue, for fundamental-mode ACEPs in lime, and for first overtone ACEPs in turquoise. In the left panel, the absolute magnitudes were computed using photometry from the source catalog and pulsation periods from \gaia. In the absence of OGLE or \gaia\ periods, we use those from the literature. The right panel shows the updated classification: dots indicate cases with available SOS photometry, and crosses represent cases where only the source catalog photometry was available. The blue, non-filled histogram at the top includes all Cepheids in clusters, regardless of their subclassification or pulsation type. The gray histogram shows the T2CEPs detected by the OGLE collaboration in the LMC \citep{2018AcA....68...89S}.  }
    \label{fig:PLR_CEP_SOURCE_PHOTOMETRY}
\end{figure*}

The left panel of Fig. \ref{fig:PLR_CEP_SOURCE_PHOTOMETRY} displays all sources from the initial sample presented in Sect. \ref{sec:preliminary_membership}. The histogram at the top compares the period distributions of the stars in our analysis with those in the LMC.  The right panel shows the sample after outliers have been removed.  The classification obtained from this analysis is included in Table \ref{Tab:Cepheids_Gc}. 

\begin{figure}
    \centering
    \includegraphics[width=\linewidth]{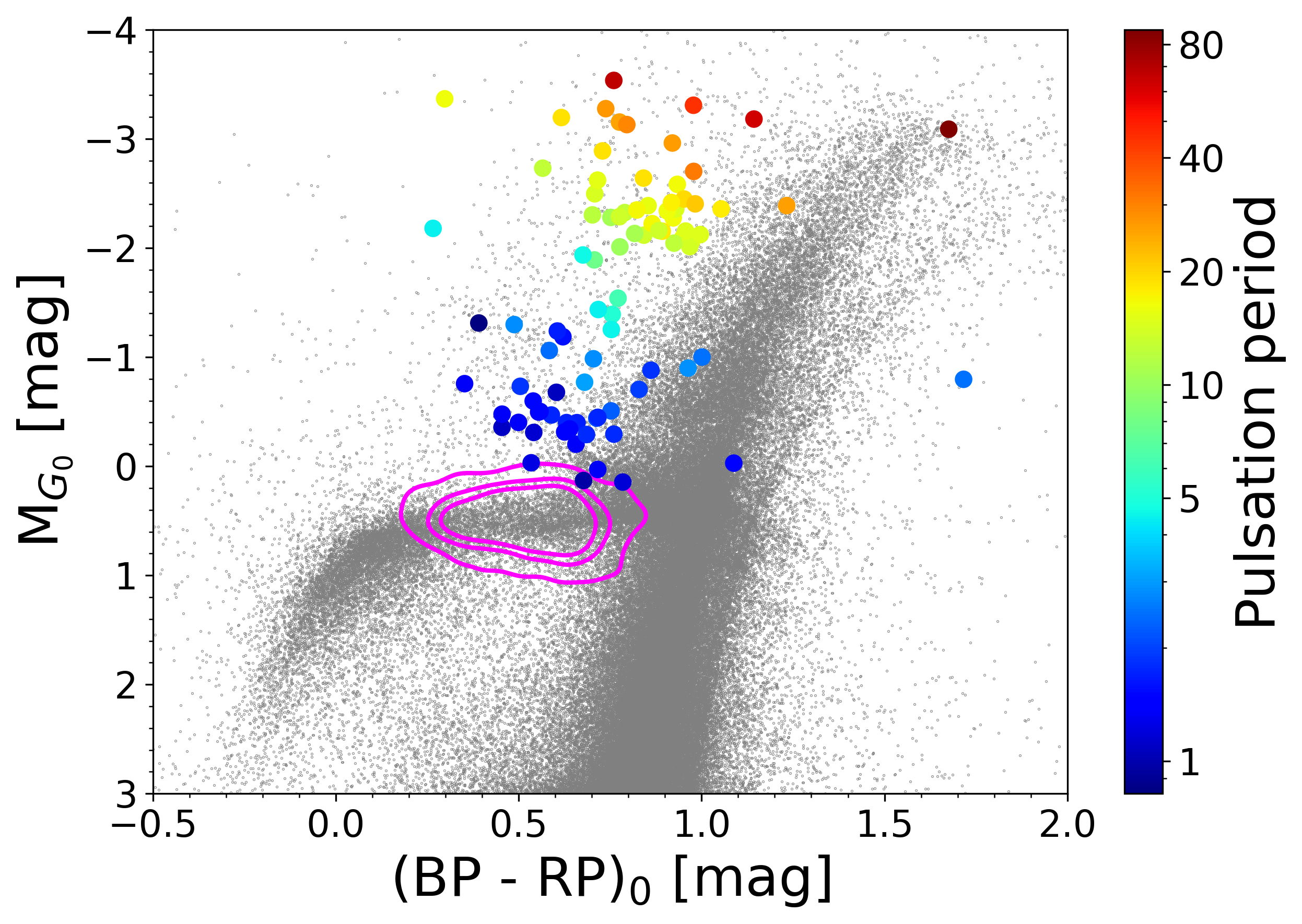}
    \caption{Color-magnitude diagram for all Cepheids and candidates in our sample. Cepheids are color coded according to their pulsation period: Cepheids with longer pulsation periods are brighter than those with shorter periods. The magenta lines indicate the 70th, 80th, and 90th percentile density contours for the sample of RRL stars from \citetalias{2024arXiv240208843C}.}  
    \label{fig:CMD_ceps}
\end{figure}

Figure \ref{fig:CMD_ceps} shows the color-magnitude diagram for all Cepheids in the final sample, with each Cepheid color coded by their pulsation period. The periods were taken from the \gaia\ SOS analysis or supplemented with values from the literature in the absence of \gaia\ data. As expected, the brightest Cepheids are those with the longest pulsation periods. Clusters that host Cepheids contain, on average, two such stars. NGC~5139 ($\omega$~Centauri) is the GC with the highest number of Cepheids, containing seven, followed by NGC~6402, with six.  In comparison, the GC with the highest number of RRLs is NGC~5272, with 236.  Furthermore, 80 of the 115 RRL-hosting GCs identified in \citetalias{2024arXiv240208843C} do not contain any Cepheids. Interestingly, we identified two GCs (NGC~6256 and NGC~6752) that host Cepheids but lack RRLs. Since both clusters are located within 10~kpc of the Sun, any RRLs should be detectable by \gaia.

\begin{figure}
    \includegraphics[width=\linewidth]{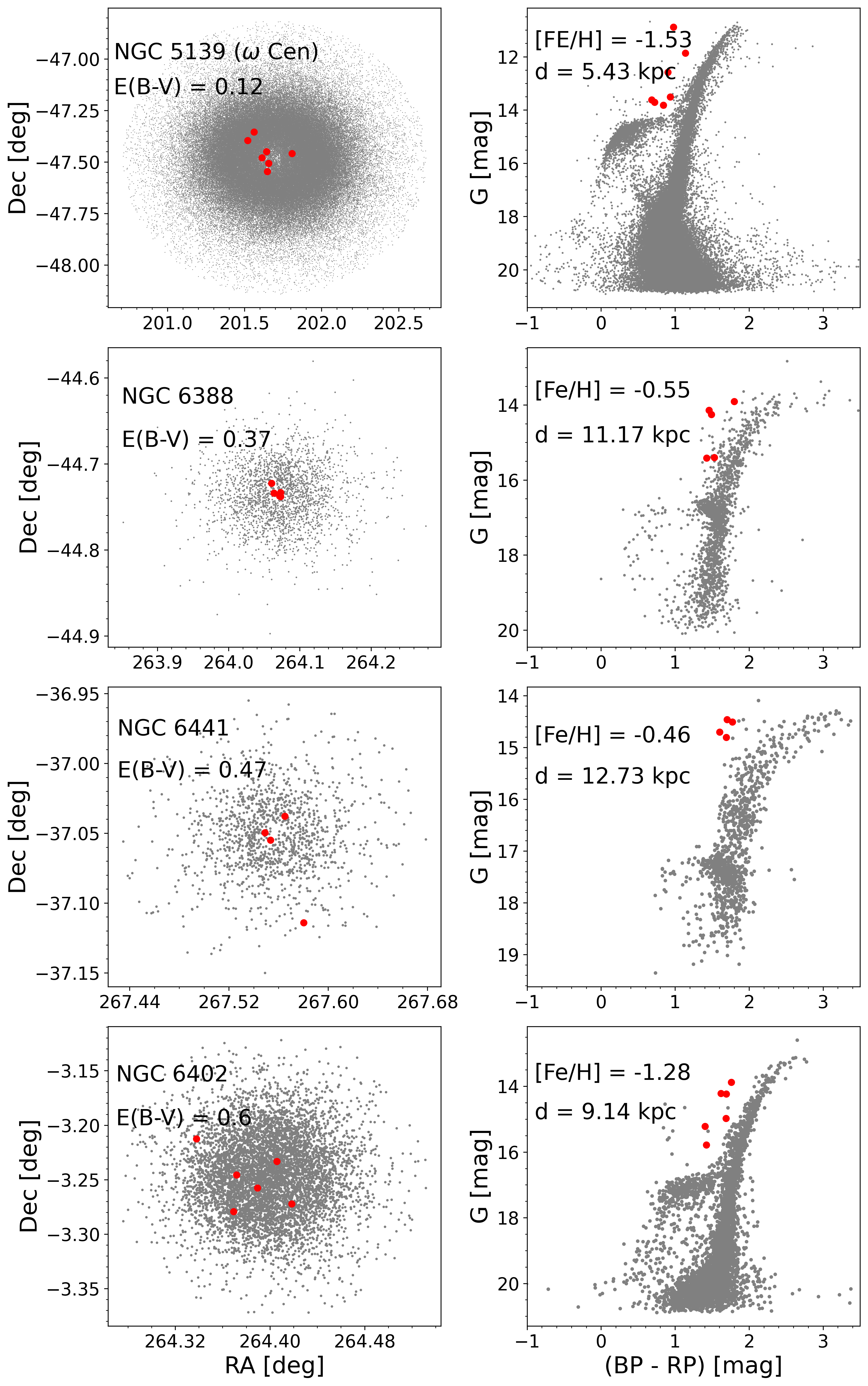}
         \caption{Clusters with the highest number of Cepheids.  Cluster members are represented with gray dots and Cepheids with red dots. The left panel illustrates the location of the Cepheids within the cluster, while the right panel shows their color-magnitude diagram. Interestingly, in the first three clusters, the Cepheids are preferentially located close to the cluster center.} 
   \label{fig:T2CEP_clusters}
\end{figure}

In clusters containing both RRLs and T2CEPs, we find an average of 19 RRLs per Cepheid. Some GCs are particularly interesting. For instance, NGC~5272~(M3) contains 236 RRLs and only one Cepheid, while NGC~6388 has 19 RRLs and five Cepheids. For the four clusters with the highest number of Cepheids, the cluster members and their corresponding color-magnitude diagram can be seen in Fig. \ref{fig:T2CEP_clusters}. For the first three clusters, the Cepheids are located near the center of the cluster. 

Given that Cepheid stars are brighter than RRLs, it is logical to assume that they would be more easily detectable. Therefore, the completeness of the Cepheid catalog would be expected to be higher than that of RRLs, but this is not the case. In \citetalias{2024arXiv240208843C}, we showed that the completeness of the \gaia\ Data Release 3 (DR3) catalogs of RRLs relative to our catalog is 77\% for the SOS and 82\% for the classifier. In contrast, the completeness of the \gaia\ Cepheid catalogs is 64\% for the SOS and 74\% for the classifier with respect to our sample.  

\begin{figure}
    \centering
    \includegraphics[width=\linewidth]{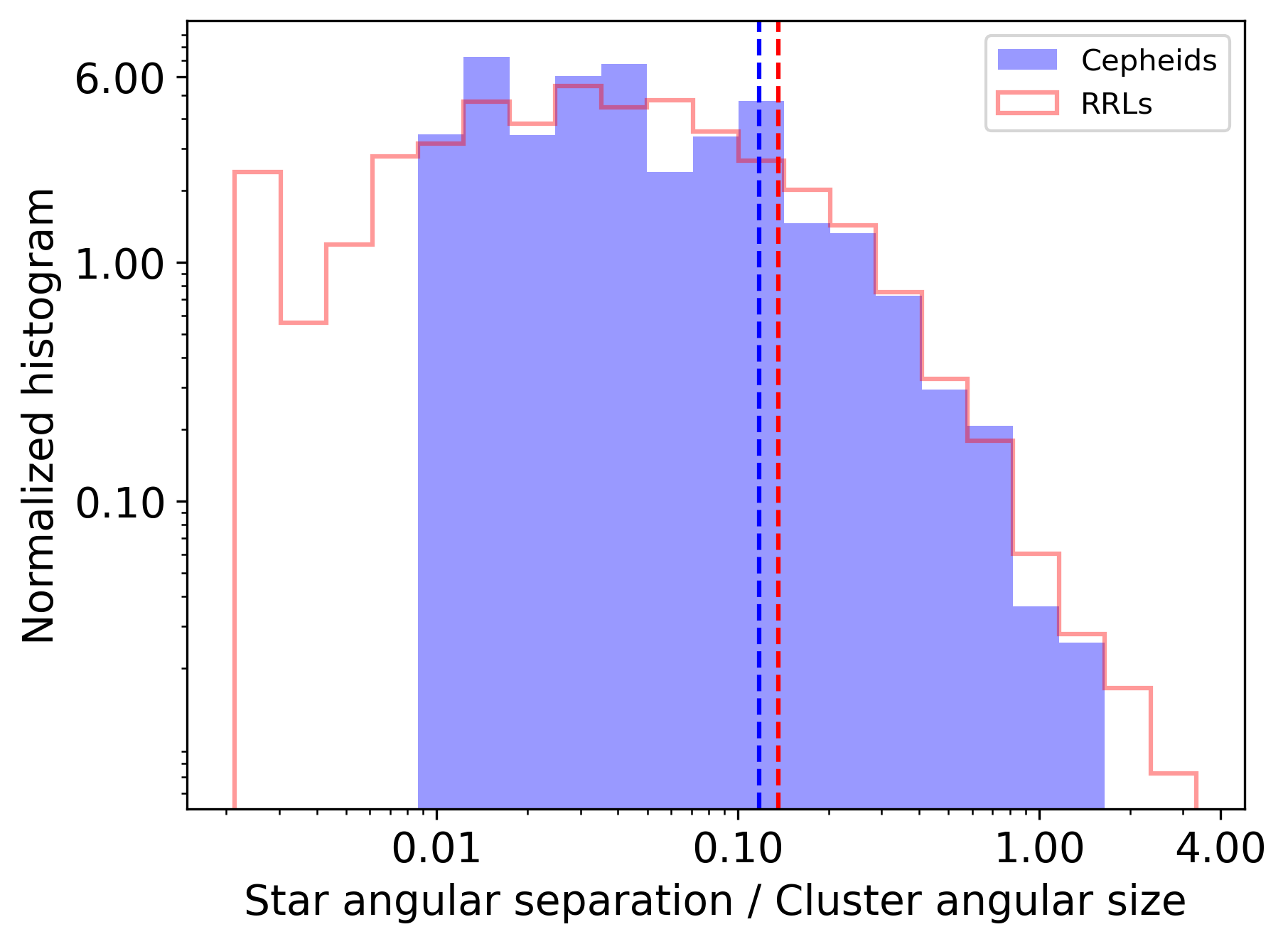}
    \caption{Normalized histogram showing the angular distance of a star from the cluster center, divided by the angular size of the host cluster.  The vertical lines correspond to the median separation for each distribution.  The histogram only shows clusters that host both Cepheids and RRLs. }
    \label{fig:separation}
\end{figure}

For clusters that host both RRLs and Cepheids, we measured the separation in parsecs between each cluster center and these stars, as shown in Fig. \ref{fig:separation}.  We did not detect any Cepheids at angular separations exceeding 1.4 times the size of the cluster, whereas 2.5\% of the RRLs, distributed across 18 GCs, lie beyond this limit. Additionally, no Cepheids were detected at separations smaller than 0.01 times the size of the cluster, but 1.7\% of the RRLs, spread across five GCs, are located in that region. However, it is important to note that the RRL sample is approximately 21 times larger than the Cepheid sample. Therefore, the absence of Cepheids in the tidal tails or near the center may be due to small sample statistics, rather than indicating a unique distribution of Cepheids within clusters.

\begin{figure}
    \centering
    \includegraphics[width=\linewidth]{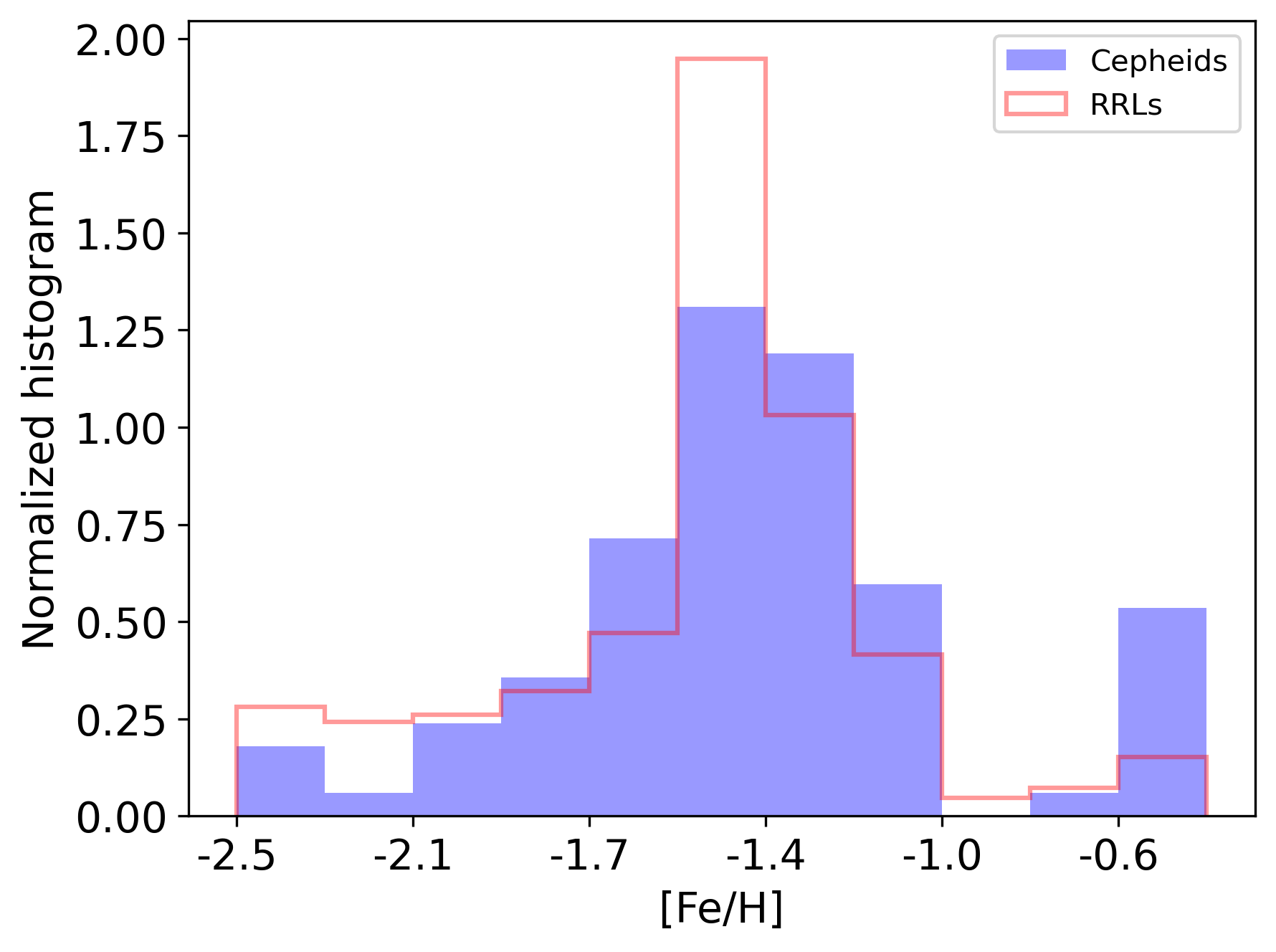}
    \caption{Normalized histogram where each Cepheid and RRL is assigned the mean [Fe/H] value of their host GC.  }
    \label{fig:fe_h}
\end{figure}

To illustrate the metallicity distribution of these stars, we assigned each Cepheid and RRL the mean [Fe/H] value of its host cluster \citep{harris2010new}. The results are shown in Fig. \ref{fig:fe_h} and include the 2824 RRLs detected in \citetalias{2024arXiv240208843C} and the 88 Cepheids detected in this study. Both populations are predominantly found in GCs with an [Fe/H] abundance of -1.5. The peak at [Fe/H]~$< -0.6$ for Cepheids corresponds to nine stars, with five in NGC~6388 and four in NGC~6441.

\section{Models versus observations}\label{sec:models}
Following the methodology of \citetalias{2024arXiv240208843C}, we calculated the edges of the instability strip (IS) for T2CEP using MESA-RSP version r23.05.1 \citep{2019ApJS..243...10P}.  We assumed a mass of $M = 0.6\,{\rm M}_\odot$, the convective parameters of Set B from \citet{2019ApJS..243...10P}, and the range of metallicities used in \citetalias{2024arXiv240208843C}, $Z = 0.0001 ,0.0003$. Four different values for the helium abundance were used, $Y = 0.220, 0.245, 0.290,$ and $0.357$.  The models were calculated in the luminosity range  $\log L/L_\odot \in \langle 1.7, 2.3 \rangle,$ with a 0.05~dex step, and an effective temperature of $T{\rm eff} \in \langle 4000, 8000 \rangle$~K in 50~K intervals. The results can be found in Table \ref{tab:models}.

To compare the location of the T2CEPs in the color-magnitude diagram with the predicted IS boundaries, we used \gaia\ SOS photometry \citep{2022arXiv220606278C}  and the classification obtained in the previous section. We excluded all ACEPs from the sample.   Once all restrictions were applied, only 52 stars remained.   Taking the sample size into account, a $\sim 2$\% change corresponds to the inclusion or exclusion of a single T2CEP within the IS boundaries. The percentage of T2CEPs within the IS boundaries for all models can be found in Table \ref{tab:models}. The models that align most closely with the observations contain 90\% of the T2CEPs. These models have a metallicity of $Z = 0.0003$ regardless of the helium abundance used ($Y = 0.220$ or $0.245)$.  If the metallicity is reduced to $Z = 0.0001$, the percentage of T2CEPs within the boundaries of the IS decreases to 88\% for the helium abundance $Y = 0.022$. It reduces to 87\% for $ Y = 0.0245$ and decreases by an additional 6\% for $Y = 0.0357$. 

\begin{figure}
    \centering
    \includegraphics[width=\linewidth]{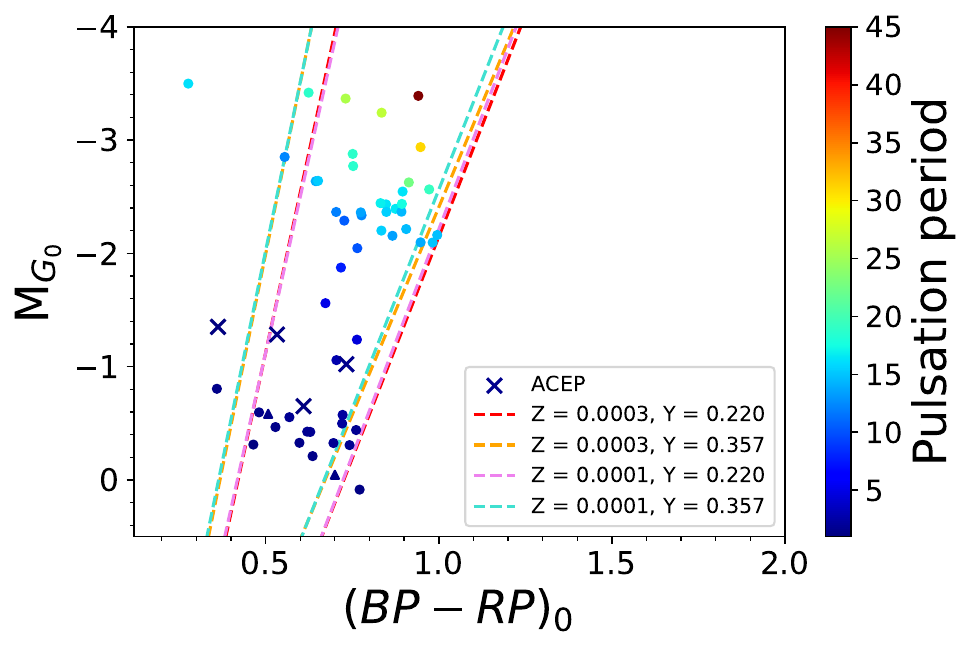}
    \caption{Comparison of the T2CEP models with the T2CEPs in our sample. The triangles represent stars pulsating in the first overtone (based on the \gaia\ classification), and the dots represent the stars pulsating in the fundamental mode. Anomalous Cepheids are marked with the symbol $\times$.  The dotted lines are used to indicate the boundaries of the IS as defined by the models. The best-fitting models for the majority of Cepheids are shown in red,  whereas those fitting the least are in turquoise. The percentage of stars within the boundaries of each model is given in Table  \ref{tab:models}.  }
    \label{fig:t2ceps_models}
\end{figure}

If $Y = 0.357$ is held constant,  models with $Z = 0.0003$ contain 83\% of all T2CEP stars, and those with $Z = 0.0001$ contain 81\%.  The limited number of T2CEPs in our sample prevents us from placing more restrictive constraints on the values of $Y$ and $Z$. The blue and red edges of the IS for T2CEPs are displayed in Fig. \ref{fig:t2ceps_models}.  

For comparison, in \citetalias{2024arXiv240208843C}, the models that best matched the observations of RRLs pulsating in the first overtone (RRc) had parameters $Z = 0.0003$, $M = 0.7 M_{\odot}$, and $Y = 0.357$, while models with $Z = 0.0003$, $M = 0.7 M_{\odot}$, and $Y = 0.220$ provided the best fit for RRLs pulsating in the fundamental mode (RRab). Therefore, our results suggest that the IS boundaries that best match T2CEPs align with the parameters associated with RRab stars, that is, $Z = 0.0003$ and $Y = 0.220$.

\section{Summary and conclusions}\label{sec:conclusions}
Starting from a sample of 21,240 unique Cepheids from multiple sources in the literature and using the \gaia\ astrometry, we performed a membership analysis for these stars in GCs, resulting in a catalog that includes 88 Cepheids and 44 candidates. Although comparable to the 105 Cepheids and 25 candidates in \citetalias{2017EPJWC.15201021C}, our sample was homogeneously determined and offers membership probabilities for all stars. Cepheids located at the centers of GCs in \citetalias{2017EPJWC.15201021C} were likely excluded from our catalog due to blending limitations in \gaia.

Cepheid candidates retain this classification because their status could not be confirmed, due to either the lack of a measured pulsation period or the absence of BP/RP photometry. The completeness of \gaia\ analyses for Cepheids with respect to our catalog is 64\% for the SOS and 74\% for the classifier. Among the stars in our sample, 78 follow the period-luminosity relations for T2CEPs. Nine stars are consistent with the relation for fundamental-mode ACEPs, and one fits the first-overtone ACEP relation.  If the classification is confirmed, this would increase the number of known ACEPs in GCs by a factor of two. 

The MESA models that provide the best fit for T2CEP have $M = 0.6\,{\rm M}_\odot$ and $Z = 0.0003$. Given the small number of T2CEPs,  these models are consistent with two different helium abundances (Y = 0.220 and 0.245). Each model includes 90\% of the T2CEPs within its boundaries.  While tighter constraints require a larger number of T2CEPs, current observations suggest that the percentage of stars within the IS boundaries decreases with increasing $Y$ or decreasing $Z$. We note that, similar to RRLs, these stars are more frequently found in clusters with an average value of [Fe/H]~$ = -1.5 $.

Increasing the sample of variable stars in GCs could lead to stronger constraints on the parameters that describe their theoretical models, provide a more detailed understanding of their formation mechanisms, and improve distance estimates for Population II stars. This last aspect will be explored in detail in the next paper of the series, where we will use the samples of RRLs and Cepheids in GCs to simultaneously calibrate their period-luminosity relations (Lengen et al., in prep.).

\begin{acknowledgements}
The authors thank the anonymous referee for their comments, which helped improve the quality of the manuscript. We also thank Christine Clement for providing us with the most recent version of her \href{https://www.astro.utoronto.ca/~cclement/cat/listngc.html}{catalog}, and we are thankful for the valuable discussions with the members of Consortium Unit Number Seven of the \gaia\ collaboration.

MC \& RIA acknowledge support from the European Research Council (ERC) under the European Union's Horizon 2020 research and innovation programme (Grant Agreement No. 947660). RIA is funded by a Swiss National Science Foundation Eccellenza Professorial Fellowship (award PCEFP2\_194638).  SD acknowledges the KKP-137523 `SeismoLab' \'Elvonal grant of the Hungarian Research, Development and Innovation Office (NKFIH).

This work has made use of data from the European Space Agency (ESA) mission {\it Gaia} (\url{https://www.cosmos.esa.int/gaia}), processed by the {\it Gaia} Data Processing and Analysis Consortium (DPAC,
\url{https://www.cosmos.esa.int/web/gaia/dpac/consortium}). Funding for the DPAC has been provided by national institutions, in particular the institutions participating in the {\it Gaia} Multilateral Agreement. 

This research has made use of NASA's Astrophysics Data System; the SIMBAD database and the VizieR catalog access tool\footnote{\url{http://cdsweb.u-strasbg.fr/}} provided by CDS, Strasbourg; Astropy\footnote{\url{http://www.astropy.org}}, a community-developed core Python package for Astronomy \citep{astropy:2013, astropy:2018}; TOPCAT\footnote{\url{http://www.star.bristol.ac.uk/~mbt/topcat/}} \citep{2005ASPC..347...29T}.

\end{acknowledgements}

\bibliographystyle{aa}
\bibliography{refs}

\begin{appendix}

\section{Tables}

The appendix includes Tables \ref{Tab:Cepheids_Gc}, \ref{Tab:cepheid_parameters}, and \ref{tab:models}.   Table \ref{Tab:Cepheids_Gc} provides details about the sample of Cepheids analyzed in Sect. \ref{sec:processing}. Table \ref{Tab:cepheid_parameters} lists the identifiers for all Cepheids in clusters from the \gaia, \citetalias{2017EPJWC.15201021C}, and OGLE catalogs, along with their assigned classifications. Finally, Table \ref{tab:models} shows the models for the blue and red edges of the instability strip for T2CEPs.

\begin{table}[h]
  \centering
  \setlength{\tabcolsep}{1.4pt}
  \begin{threeparttable}
    \caption{Cepheids in globular clusters.}
    \label{Tab:Cepheids_Gc}
    \begin{tabular}{llllcccccccccc}\toprule
      \gaia\ DR3 source id & ra [deg] & dec [deg] & Cluster & likelihood  & prior & posterior & Final  & Classification & Candidate  & Period [d] & SOS \\ \midrule
      899283619691731328   & 114.530019  & 38.912819  & NGC 2419 & 0.963  & 1   & 0.963    & True   & ACEP F        & False      & 1.579     & False \\
      1328057184175359488  & 250.443656  & 36.457716  & NGC 6205 & 0.973  & 1   & 0.973    & True   & T2CEP         & False      & 1.459     & True \\
      1328057867076964864  & 250.399567  & 36.463465  & NGC 6205 & 0.999  & 1   & 0.999    & True   & T2CEP         & False      & 5.111     & True \\
      ... & ... & ... & ... & ... & ... & ... & ... & ... & ... & ... & ... \\
      \bottomrule
    \end{tabular}
    \begin{tablenotes}[para]
      \footnotesize
      {\bfseries Notes.} The complete version of this table is available at the CDS. If the likelihood for a given source is empty, it indicates that parallax and proper motion data are missing from the \gaia\ DR3 catalog, these stars were identified as cluster members by \citetalias{2017EPJWC.15201021C}.  The "Final" column specifies whether the star is part of the sample presented at the end of Sect. \ref{sec:processing}, while the "Classification" column lists the classification assigned to each star in this paper. The "Candidate" column identifies Cepheids for which membership could not be confirmed due to the absence of a measured pulsation period or time-series photometry. The column 'Period' represents the pulsation period used for classification, while the column 'SOS' shows whether SOS photometry is available in the Gaia archive.
    \end{tablenotes}
  \end{threeparttable}
  
\end{table}

\begin{table}[h]
\setlength{\tabcolsep}{12pt}
\begin{threeparttable}
  \caption{Cepheid identifiers along with their classifications in other catalogs.}
  \label{Tab:cepheid_parameters}
  \begin{tabular}{llcccccc}\toprule
                       &         &        &                     &  \multicolumn{4}{c}{Classification}       \\ \cline{5-8}
    \gaia\ DR3 source id & Cluster & C17 ID & OGLE ID            &  SOS      & C17       & Classifier  & G23 \\ \toprule
    899283619691731328   & NGC 2419 & 18     &                    &           & CWA      &             & RRAB \\
    1328057184175359488  & NGC 6205 & 1      &                    & T2CEP     & CW       & CEP         & BLHER \\
    1328057867076964864  & NGC 6205 & 2      &                    & T2CEP     & CW       & CEP         & CW \\
    ...                 & ...     & ...    & ...                & ...       & ...      & ...         & ...  \\
    \bottomrule
  \end{tabular}
  \begin{tablenotes}[para]
    \footnotesize
 {\bfseries Notes.} The complete version of this table is available at the CDS. This table lists the identifiers assigned to all Cepheids in clusters (and candidates) from the Gaia DR3, C17, and OGLE IV catalogs. Cepheids from Table~\ref{Tab:Cepheids_Gc}, which were excluded as cluster members, are not included here.
  \end{tablenotes}
\end{threeparttable}
\end{table}

\begin{table}[h!]
\caption{Theoretical blue and red edges of the instability strip for T2CEPs in the (G,BP-RP) plane.}
\centering
\scalebox{0.95}{
\begin{tabular}{c c c c c}
\hline
 Blue edge & Red edge & Parameters & Percentage of T2CEPs inside the IS edges\\
\hline
-14.14$(BP-RP-0.5)$-1.12&-7.83$(BP-RP-0.5)$+1.77& $Z = 0.0003$, $Y = $0.220 & $90\%$\\
-13.76$(BP-RP-0.5)$-1.23&-7.27$(BP-RP-0.5)$+1.55& $Z = 0.0003$, $Y = $0.245 & $90\%$\\
-14.29$(BP-RP-0.5)$-1.54&-7.54$(BP-RP-0.5)$+1.55& $Z = 0.0003$, $Y = $0.290 & $88\%$\\
-15.12$(BP-RP-0.5)$-1.99&-7.33$(BP-RP-0.5)$+1.26& $Z = 0.0003$, $Y = $0.357 & $83\%$\\
\hline
-13.76$(BP-RP-0.5)$-1.12&-8.00$(BP-RP-0.5)$+1.79& $Z = 0.0001$, $Y = $0.220 & $88\%$\\
-12.90$(BP-RP-0.5)$-1.21&-8.21$(BP-RP-0.5)$+1.71& $Z = 0.0001$, $Y = $0.245 & $87\%$\\
-14.22$(BP-RP-0.5)$-1.54&-7.35$(BP-RP-0.5)$+1.46& $Z = 0.0001$, $ Y= $0.290 & $85\%$\\
-14.89$(BP-RP-0.5)$-2.02&-7.74$(BP-RP-0.5)$+1.31& $Z = 0.0001$, $Y = $0.357 & $81\%$\\
\hline

\hline
\end{tabular}}
\tablefoot{All models were computed assuming $M = 0.6 M_{\odot}$.  }
\label{tab:models}
\end{table}

\end{appendix}

\end{document}